# Constructing the Molecular Tree of Life using Assembly Theory and Mass Spectrometry

Amit Kahana[1][†], Alasdair MacLeod[1][†], Hessam Mehr[1][†], Abhishek Sharma[1], Emma Carrick[1], Michael Jirasek[1], Sara Walker[2,3] and Leroy Cronin[1]*

[1]School of Chemistry, University of Glasgow, Glasgow, G12 8QQ, UK.
[2]BEYOND Center for Fundamental Concepts in Science, Arizona State University, Tempe, AZ, USA
[3]School of Earth and Space Exploration, Arizona State University, Tempe, AZ, USA
*Email: Lee.Cronin@glasgow.ac.uk  [†]These authors contributed equally.

**All life forms share a common biochemical ancestry, yet sequence-based phylogenetics falters when DNA is scarce or absent. We introduce a biochemistry-agnostic strategy that reconstructs evolutionary relationships directly from molecular ensembles by coupling tandem mass spectrometry (MS) with Assembly Theory (AT). AT quantifies shared structural information through each sample's joint assembly space, facilitating the inference of molecular complexity without identifying individual compounds. By analysing 74 biotic and abiotic samples, we detected >9,000 analytes and >6,000 molecular fragments which we used to build joint assembly spaces, and these allowed us to reliably distinguish living from non-living samples, group organisms within their tree, and produce a phylogenetic tree similar with genomic models. Further experiments revealed that in multicellular organisms, AT clustered disparate tissues by species rather than phenotype, and in *Escherichia coli* it resolved colony lineages across several culturing generations in the lab under timescales elusive to genomic methods. These results establish AT-guided MS as a universal, sequence-independent framework for causal molecular inference across biochemical systems.**

Tracing evolutionary kinships has long been central to biology. Pragmatic early taxonomies—grouping plants by medicinal use or animals by gross morphology—exposed the need for a unified framework (*1*). Standardized hierarchical systems emerged in the 18th century (*2*) and, by the 19th, comparative phenotypic analyses had revealed that disparate taxa share common ancestry (*3*). These insights birthed phylogenetics and culminated in today's "tree of life," a universal model of evolutionary relationships (*4–6*). While the foundational principles of evolution remain intact, the "tree of life" has been continually sharpened by advances in genetics, molecular biology and bioinformatics, yielding ever-finer portraits of life's history. High-throughput, low-cost sequencing has entrenched genomic data as the primary—often exclusive—currency for linking organisms across time, and decades of biochemical insight have spawned powerful algorithms that mine these sequences to elucidate evolutionary trajectories (*5–8*). Yet a sequence-centric lens leaves sizeable blind spots that are difficult to overcome. For example rapidly evolving and biochemically-unique viruses often elude secure rooting in a phylogenetic model (*9*); unculturable microbes supply fragmentary, sparse and often unannotated genetic reads, leaving much of Earth's extant biodiversity unmapped (*10–12*); and any protobiological pre-LUCA entities remain unreachable, as they plausibly did not contain genetic information. This means that putative molecular fossils from this deep past cannot be situated on a tree anchored solely in DNA (*13*, *14*, *15*).

In the field of chemotaxonomy, organisms are identified and classified by their biochemical fingerprints and, aided by quantitative targeted metabolomics, lineages have successfully been resolved for plant, fungal, insect, and microbiome systems among others (*16–24*). However, these models rely on the detection and quantification of predefined biomarkers (often large polymeric metabolites) by crossing data with reference spectral libraries. This approach is constrained by the availability and the quality of such libraries, covering only a fraction of the metabolome, and therefore vast suites of unannotated molecules detected in experiments are not addressed (*20*, *21*, *24–26*). As a result, chemotaxonomic approaches remain tethered to known chemistry, leaving life forms with unfamiliar biochemistry beyond reach.

We hypothesize that Assembly Theory (AT) in conjunction with mass spectrometry (MS) measurements could be used to go beyond methods requiring prior fingerprinting of molecular species, allowing mapping relationships of evolving systems where sequence data is sparse or non-existent (*27*). According to AT, every molecule has an inherent complexity (Molecular Assembly, MA) corresponding to the shortest construction pathway of its molecular graph. Central to AT is the concept of an assembly space, composed of a molecule's precursors along its construction pathway, where each fragment is recursively built up from simpler building blocks. The smallest possible cardinality of the assembly space is indicated by the target molecule's MA, quantifying the minimal number of constraints necessary for the construction of the molecule.

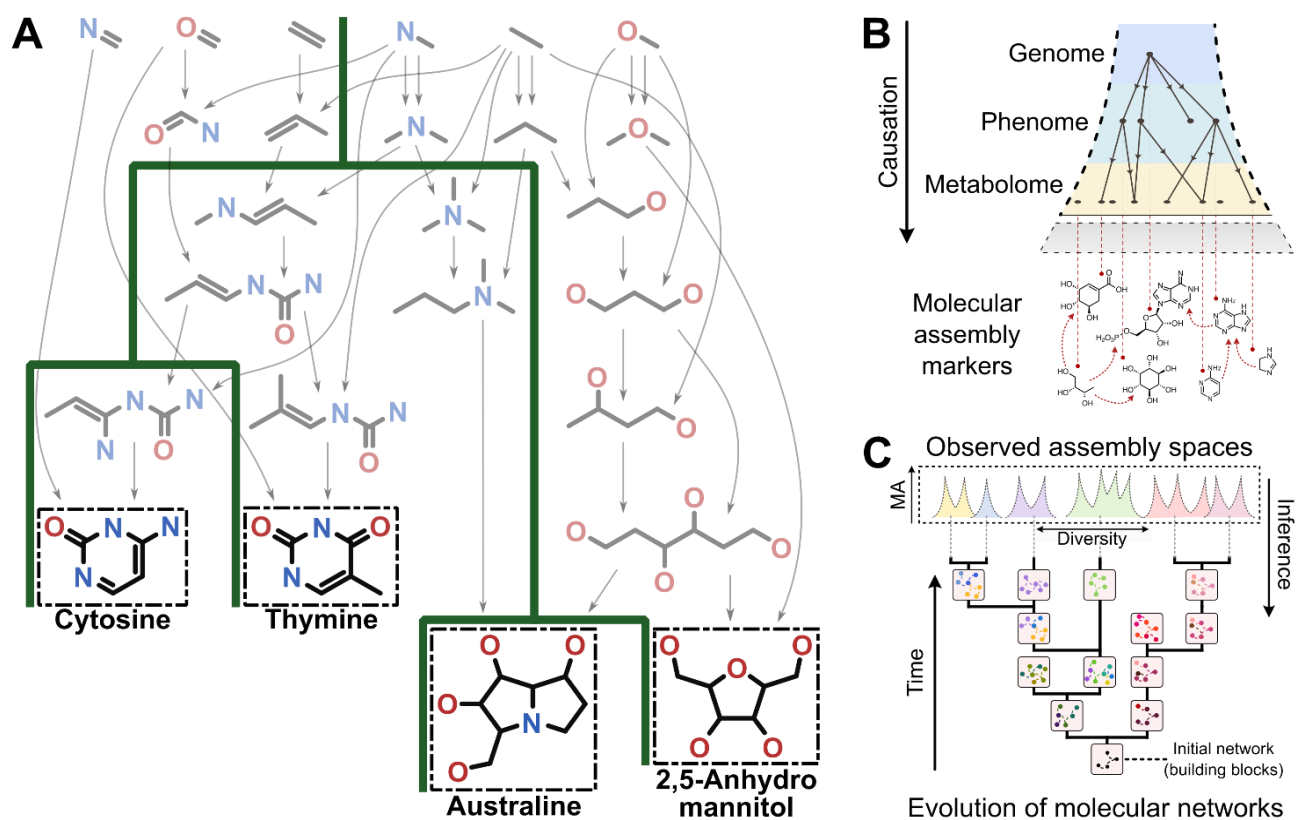

**Figure 1.** Joint Assembly Spaces of molecules and biological systems undergoing evolution. **A)** The joint assembly space (JAS) of thymine, cytosine, australine and 2,5-anhydromannitol, consisting of their combined assembly pathways to give the shortest sequence of joining operations for their simultaneous construction. All individual MAs of the present molecules are 6 (except 2,5-anhydromannitol with MA 5), and their Joint MA is 19. The green tree-like structure represents the causal relationships between the present molecules manifested in their assembly pathways. **B)** The cellular composome and its informational architecture. The flow of information is portrayed, from the genome to the phenome (e.g. proteins and ribozymes) and the metabolome. Observable analytes from these levels serve as molecular assembly markers, which give information about the assembly space underlying the cellular composome, and which can inform about the unique identity and state of the cell. **C)** An assembly representation of evolution as a tree of molecular networks, from an initial network composed of an assortment of building blocks towards more defined and complex organismal networks. The molecular networks evolve over time and become more distinctly associated with recent clades. The observed assembly spaces of molecular networks are shown at the top, defined by the diversity and complexity of their molecular structures and used for phylogenetic inference. Colors indicate association to specific phylogenetic grouping, and their idiosyncratic traces through their inferred evolutionary past.

It has been demonstrated that MA can be experimentally measured using mass spectrometry and other analytical techniques (*28*), and previous work empirically verified that organic molecules with high MA values are indicative of life, presenting MA as a genome-agnostic method for life detection (*27*). The concept of assembly space can be generalized to apply to a set of distinct molecules, more befitting biological samples comprising molecular networks, where the simultaneous construction pathway for the entire set of molecules together is called a Joint Assembly Space (JAS) (**Figure 1A**). The shortest pathway in a JAS represents the minimal set of constraints required to construct a given ensemble of molecules simultaneously (*29*), and the number of steps along this pathway is termed Joint Molecular Assembly (Joint MA). The JAS is comprised of the target objects measured at a single point in time, and of the contingent objects along their assembly pathways. Within the framework of a generalized AT, these respectively correspond to the Assembly Observed, the space of all objects that are observationally confirmed, and the Assembly Contingent, the space of all contingent sub-objects generated from assembly pathways that are governed by historical contingency (*30*). The formulation of a JAS can be used as a tool to reveal contingent relationships between individual molecules or entire chemical systems, particularly important for cases where the exact chemical reaction pathways are very complex or the temporal formation histories of molecules are unknown, as these are unnecessary information for the JAS. Moreover, the JAS does not have explicit knowledge of geological time; however, for a set of complex molecules it does encode the underlying causal relationships that have a necessary temporal ordering for the production of a set of molecules, providing causal constraints on any process that might co-construct those molecules, including evolutionary processes.

Incorporating complexity considerations into molecular analyses could provide further insights into the informational architecture that governs biological life-forms (*30*). A biological cell is a self-constructing molecular network, which stays largely at compositional homeostasis (composome, a quasi-stationary state) (*14*, *31*) through intricate molecular interactions, an evident result of selection. Evolution facilitates the transformation of these networks over time, allowing exploration of new molecular compositions (Assembly Observed), and their underlying assembly spaces (Assembly Contingent) that lead to molecular and functional novelty (*30*). It has been shown that higher levels in the informational organization of the cell, such as the phenome and the genome, can be elucidated with a bottom-up inference based on idiosyncratic signatures in the compositional metabolome (*25, 32*), see **Figure 1B**. In this light, the metabolome may be regarded as an expansive informational arena describing the historical contingency of the cell (*33*). We therefore hypothesize that genome-agnostic molecular information, coupled with AT, may allow the inference of evolutionary trajectories beyond pure compression analyses (*34*). This would manifest in the contingent transformation of the assembly spaces of cellular molecular networks from common ancestors to current day species (**Figure 1C**). We note that since the assembly spaces do not pertain to realized metabolic pathways, it may provide a baseline for more advanced evolutionary models as it does not require specific biochemical information about the examined samples.

In this study, we propose a generalised method of inferring the JAS of the cellular composome and demonstrate that it applies not only to life detection (as in prior work) but also to life classification, reflecting the causal relationships between species as complex chemical networks. We demonstrate the utility of this approach both over longer term evolutionary trajectories, and short-term phenotypic trajectories leading to divergence in populations where no detectable genomic evolution has yet sufficiently occurred. Using AT, we were able to generate a phylogenetic tree of life - using only mass spectrometry data of small metabolites - that conforms with current biological knowledge, offering a comparable alternative to other phylogenetic methods that does not require prior molecular fingerprinting (**Figure 2**).

Further, we show the capacity of our assembly-based methodology to adhere to phylogenetic, rather than phenotypic, tree clustering for molecularly-heterogeneous multicellular organisms. Lastly we apply our AT methodology to deduce the genealogy of recursive culturing of bacterial colonies, demonstrating its ability to reveal very short-term historical contingency among low-genomic-variation samples of the same species. Our study reveals that molecular networks of biological cells carry information about their long-term evolutionary past as well as short term phenotypic history, which can be deciphered and employed as a universal apparatus for the detection and phylogenetic classification of life.

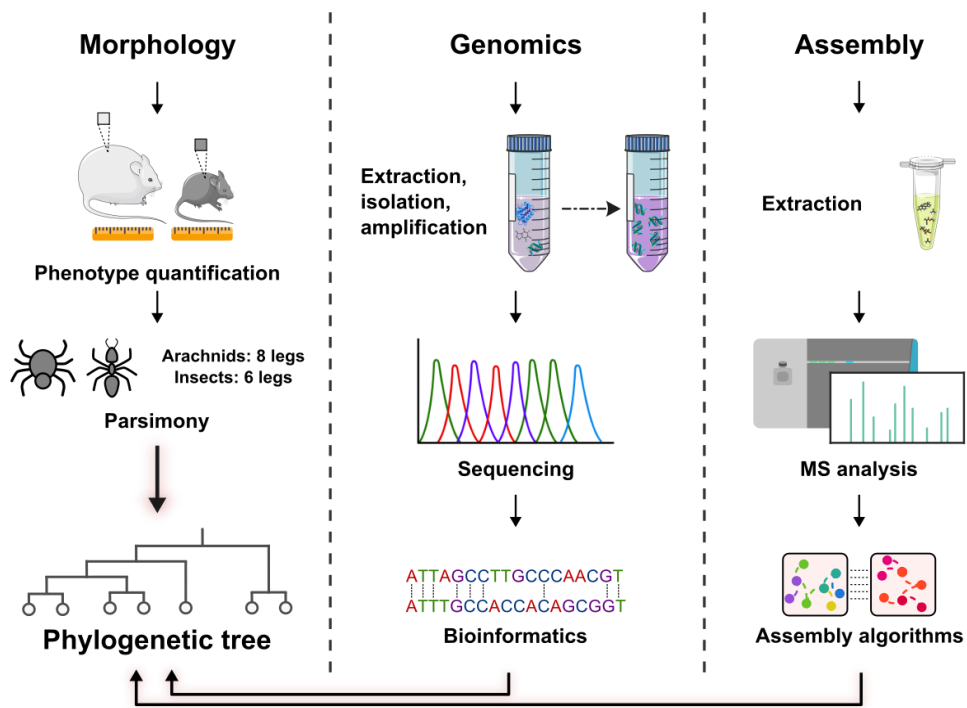

**Figure 2.** Comparison between different phylogenetic inference methods resulting in a phylogenetic tree, including differentiation according to morphology (left), genomics (center) and assembly (right, this work).

**Results and Discussion**

*Fingerprinting idiosyncratic MS-derived profiles of samples*

The tandem-MS analytical process presented here allowed us to gather information about the analytes residing in the organismal samples ($MS^1$) and their observed structural fragments ($MS^2$). Even though the analytes were not structurally elucidated, we hypothesized that non-targeted $MS^1$ and $MS^2$ information could be sufficient for species classification. We analysed 74 samples from diverse biotic and abiotic sources, resulting in 9262 unique MS1 analytes and 6755 unique MS2 fragments across the dataset. Data exploration showed that this is particularly true for differentiating between living and non-living samples – analysis of inorganic samples revealed distinctly less intense peaks (**Figure**

**3A**) and fewer analytes altogether (**Figure 3B**). We approached each set of analytes as a fingerprint of a sample's molecular network, demonstrating unique idiosyncratic features that allow its potential association to a group or clade (**Figure 3C**).

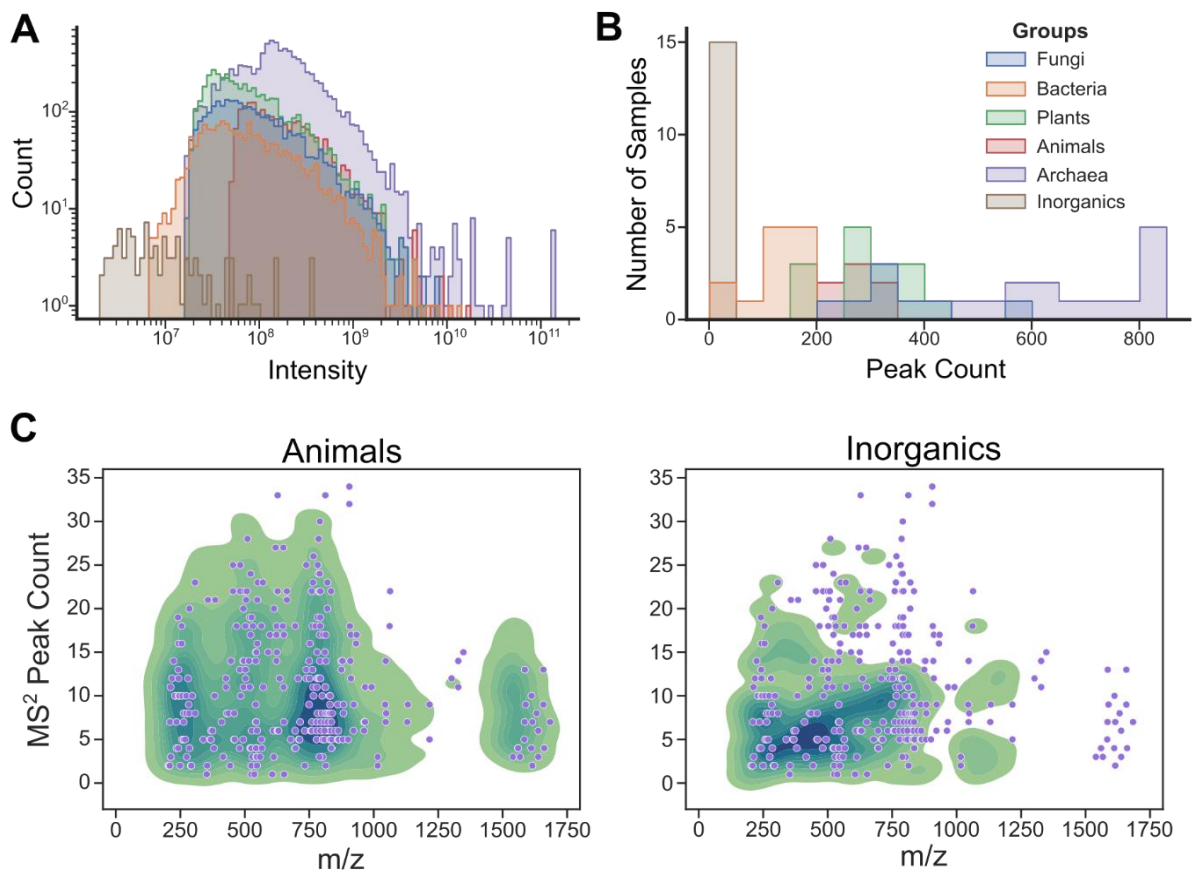

**Figure 3.** Fingerprinting molecular networks. **A)** Histograms of the intensity of peaks detected for all samples within the labeled groups. **B)** Histograms of the number of peaks detected for samples within each labeled group. **C)** Analytes from a Cod sample (purple) layed over Kernel Density Estimate (KDE) maps of the Animals (left) and Inorganics (minerals, right) groups (green), according to the m/z and number of $MS^2$ peaks of their associated analytes. The KDE bandwidth estimation was conducted with a scaling factor of 0.55. Scatterplots and KDE maps for all samples and groups can be found in the SI section 6.

To validate our experimental approach, we tried to match the detected analytes from the E. coli sample to known metabolites with reference or predicted spectra (see methods). Most detected molecules do not map to known or predicted spectra of E. coli metabolites, a result that generally matches other nontargeted biomolecular screening due to high chemical diversity and elucidation difficulties of metabolites (*35*). The wealth of analytes may correspond to an underground metabolism, encompassing reactions that are not genomically encoded and thus not readily investigated and recorded (*26*). Nevertheless, the results reveal several universal biomolecules with high certainty (such as nucleobases, amino acids, sugars, and organic cofactors) as well as more specific bacterial lipids with lesser certainty (see SI). This reaffirms our analytical procedure is without particular bias and general enough for the detection of diverse analyte chemistries across a wide mass-range, allowing faithful metabolic mapping of organismal samples.

*Genome-agnostic mapping of phylogenetic relations between organisms*

Within the AT framework, the detected $MS^1$ analytes represent the Assembly Observed of the examined organismal samples, since these are direct measurements of real molecular components of the cell. Further, the sets of $MS^2$ molecular fragments are the result of empirical deconstruction of these detected precursor analytes. They in turn represent the extended Assembly Contingent of the samples, since they interrogate the space which comprise all possible pathways to assemble the sets of detected molecules. By sampling the extended Assembly Contingent of the organismal samples, we can construct possible assembly pathways of different lengths, with the fragments acting as contingent sub-objects within the space. The space of MS-based pathway construction of all molecules within an organismal sample is therefore comparable to a JAS (*29*), which refers to the combined assembly steps and substructures included in the graphical assembly process of an examined set of target objects. Deeper levels of tandem MS correspond to join operations between smaller fragments, which are joined together in higher MS levels culminating at the original molecular precursor, emulating the JAS hierarchical structure (**Figure 4A**). This comparison is especially potent considering all metabolites in a sample originate in a shared metabolic system, an elaborate network of interconnected pathways expanding from a shared set of basic building blocks. We therefore consider the MS-derived construction space as an experimentally-derived joint assembly space and set to investigate its causal relationships using AT analytical tools.

To uncover the underlying phylogenetic architecture of the samples, we developed a numeric model based on the Recursive MA algorithm (*28*) that operates on hierarchical tandem MS data to constrain and estimate MA values for target molecules. By employing the Recursive MA algorithm, we generated joint assembly spaces for each individual sample in our dataset, recursively constructing their composite analytes from tandem MS data and pooling together their shortest detected construction pathways. This was conducted without elucidating molecular structures, instead directly operating on mass information of detected $MS^1$ analytes and their $MS^2$ fragments. We then calculated an approximated Joint MA values for discrete organismal samples, describing the estimated level of complexity of their molecular ensemble (see SI). Joint MA could be regarded as the information content of species, which are estimated to be $10^4$-$10^5$ for our samples, corresponding to the number of steps required to jointly construct their detected analytes from elementary building blocks. Though Joint MA values are often lower than the number of base-pairs in a genome, it holds more descriptive power due to the higher degree of selection required to modify the assembly space.

Additionally, Joint MA is sensitive to evolutionary transformations that may not be genomically-encoded, such as epigenetic-based variation in gene expressions. By comparing the approximated Joint MA of individual and groups of samples, we were able to estimate the Joint Assembly Overlap (JAO) of the examined samples (see methods). This metric describes the relatedness of species in the dataset based on the similarity of the shortest construction pathways calculated for their MS-based JAS. This analysis does not rely on any prior molecular information of the detected analytes, thus enabling agnostic phylogenetic inference. The interrelated evolutionary trajectories of species from shared ancestors explain why there is a considerable degree of shared $MS^2$ fragments among different molecules within and across samples, as it is unlikely that samples will possess similar molecular species by chance given the enormity of chemical space (*36*). Indeed, a larger number of mostly lower-mass fragments occur in multiple samples, with the probability of shared fragments among samples diminishing at higher fragment masses irrespective of their utility (**Figure 4B**). This observation illustrates the prevalence of shared fragments among organismal samples, supporting the applicability of assembly theory for elucidating evolutionary trajectories through the construction of joint assembly spaces.

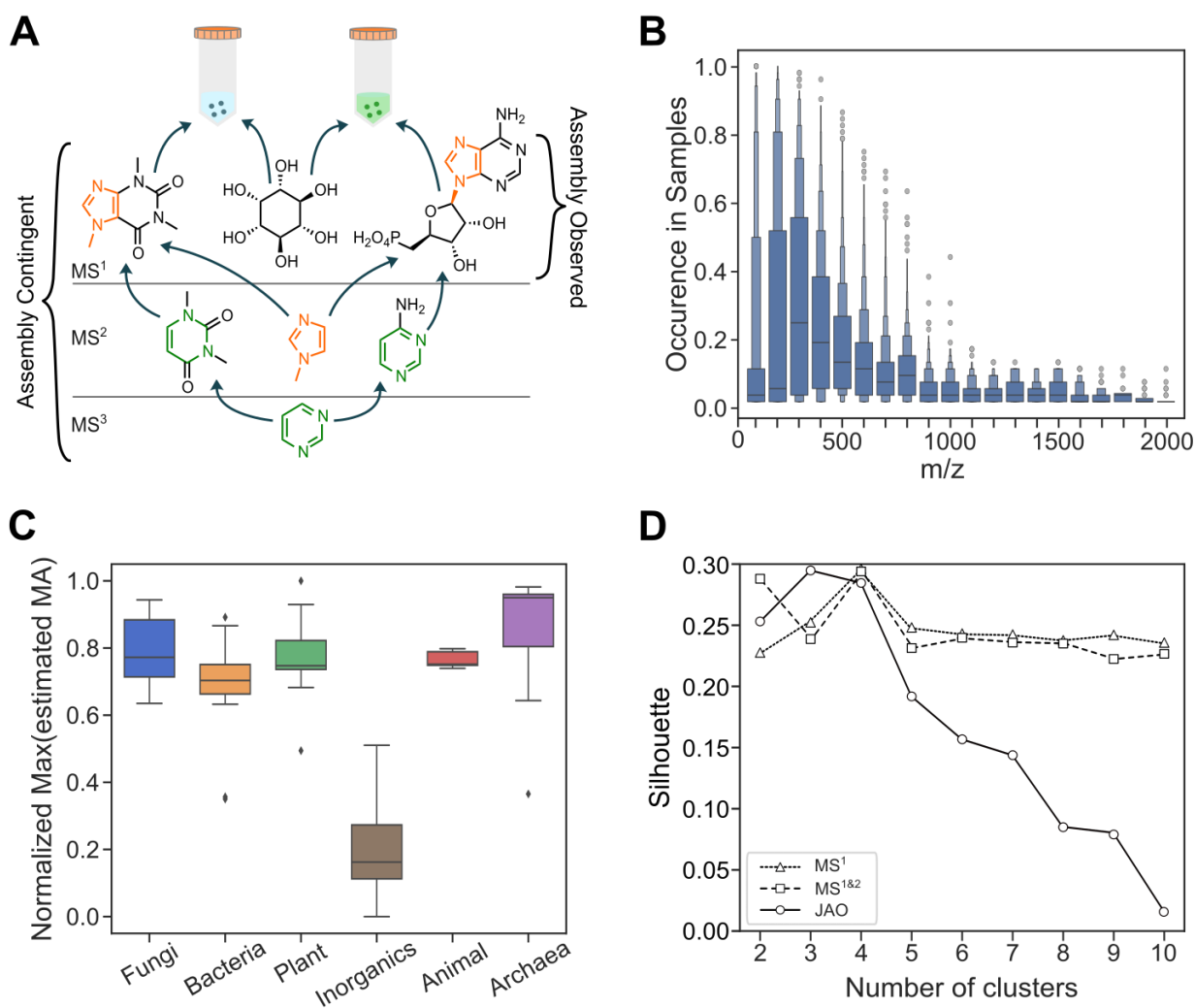

**Figure 4.** Assembly-based analyses and classification of composite MS information. **A)** Schematic representation of the MS-based Joint Fragment Assembly Space of two multimolecular samples with shared metabolites. The scheme represents an inferred pathway across the MS levels, an emulation of the results obtained with the Recursive MA algorithm. **B)**. Boxenplots of all $MS^2$ fragments observed in the analysed samples, showing their m/z and their occurrence in the samples dataset. Each box represents a bin of 100 Da. **C)** Boxplot of the maximal estimated MA of each sample for each group, calculated with Recursive MA and normalized across the set. **D)** Spectral clustering over samples using different phylogenetic methods ($MS^1$ Overlap, $MS^{1\&2}$ Overlap and JAO) for cladal group classification, presenting Silhouette scores for each number of clusters in the model. The Silhouette score indicates how similar the objects within the clusters are to each other compared to objects in other clusters, with high values indicating better clustering scheme.

*Biogenicity determination and cladal group classification with assembly spaces*

To investigate the capacity of AT to correctly discriminate between defined groups of samples, we employed it for two discrete tasks. First, we attempted to distinguish between biological and non-biological samples. We used two classification methods: a supervised leave-one-out cross validation approach, and an unsupervised spectral clustering approach (see methods). We compared the JAO metric with two simpler metrics based on AT. The first is an overlap of the detected $MS^1$ analytes, a fingerprinting approach of biomarkers which span the Assembly Observed of the samples, using a heuristic Otsuka-Ochiai coefficient (OOC) akin to cosine similarity (see methods) (*37*). The second metric is the OOC overlap of the $MS^1$ and $MS^2$ data of the samples combined which span their measured Assembly Contingent. $MS^{1\&2}$ Overlap is pathways-independent as it spans possible contingent objects without join operations, fundamentally different to the JAO approach. All metrics presented comparably successful prediction results, easily determining the biogenicity of samples in both targeted and non-targeted classification methods with very few errors (see SI). Notably, blind biogenicity determination could be reliably achieved beyond relational clustering by looking at the MA of molecules detected in samples (*27*), and this can be clearly seen in our dataset (**Figure 4C**).

Further, we tested whether we can correctly annotate biological samples as belonging to different predefined cladal groups, including Animalia, Archaea, Bacteria, Fungi and Plantae. In the supervised cross-validation approach, all metrics portrayed excellent classifications for very few errors. In the unsupervised clustering method, both the $MS^1$ and $MS^{1\&2}$ Overlap approaches showed high and largely unchanged clustering accuracy with varying number of clusters (**Figure 4D**). In contrast, the JAO approach presented a significant decline in clustering accuracy with increasing number of clusters, which is possibly more appropriate considering the hierarchical nature of the data. In general, all phylogenetic algorithms depicted logical decision-making in the clustering process that follows known taxonomical grouping and chronology (see SI). Given these results, it is arguable that the employed controls ($MS^1$ and $MS^{1\&2}$ Overlaps) approximate the pathway-dependent assembly approach (JAO), which captures better the causal hierarchical structure of living systems.

*Algorithmic construction of the tree of life with assembly theory*

Contrary to the classifiers presented in the last section, phylogenetic inference does not rely on database comparison or predefined group annotation. It is often performed without *a priori* taxonomical information, and requires more complete datasets, as it is designed to trace evolutionary trajectories of species across time and their interconnectedness – how they relate to each other. Using our phylogenetic pipeline (**Figure 5A**), we generated a context-free tree of life model based on MS information instructed by AT. We used the Recursive MA algorithm to approximate the joint assembly spaces of individual and pairs of samples, calculated the JAO between them and used this information to construct a phylogenetic tree with a simple WPGMA algorithm. The resultant tree generally displayed rational clustering decisions, closely reflecting evolutionary trends that appear in genome-based phylogenetic studies, such as the attribution of samples to specific cladal groups and the order in which these clades cluster on the tree (**Figure 5B**). Even though the model was generated through a simplistic proof-of-concept pipeline, the fact that it performed so well suggests that an assembly-based approach may be helpful for an accurate MS-based phylogenetic configuration.

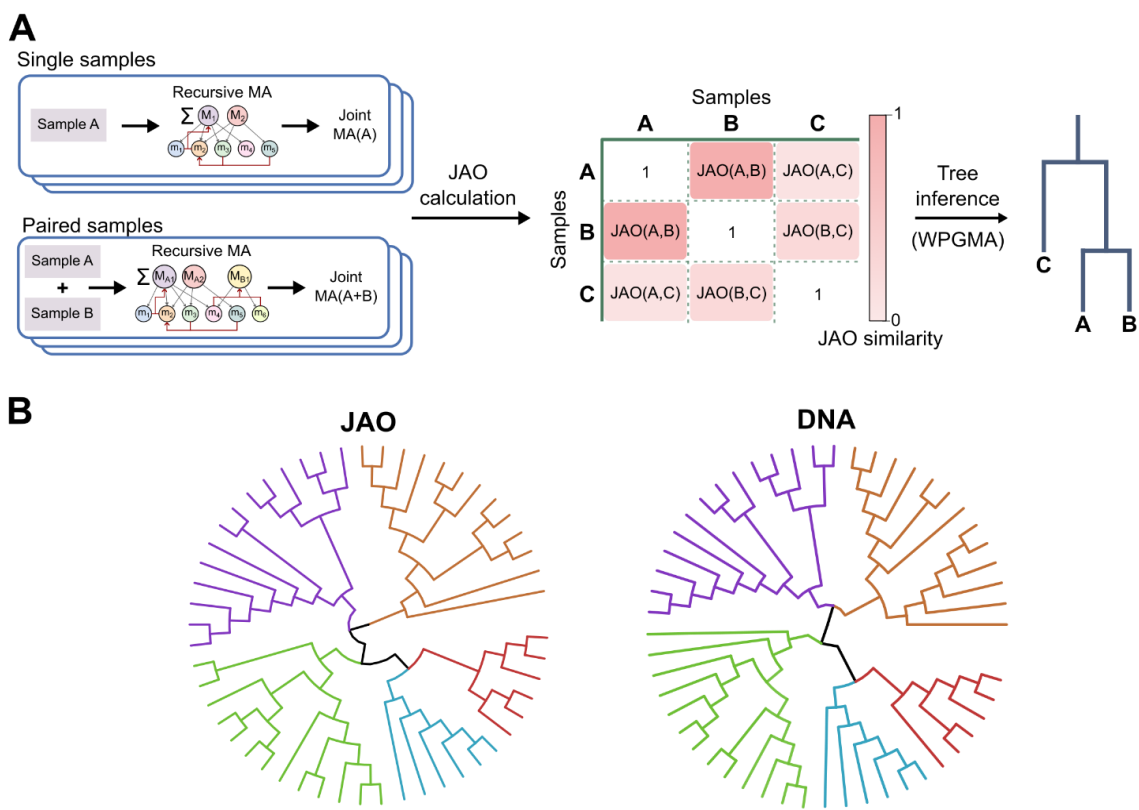

**Figure 5.** Phylogenetic inference using Assembly Theory. **A)** Algorithmic pipeline for the generation of a phylogenetic tree from MS data using Joint Assembly. Recursive MA is employed to calculate approximated Joint MA indices for single and paired samples, which are used in JAO calculations (see methods). The JAO matrix represents the overlap of joint assembly spaces of samples, which is then transformed into a tree through an inference algorithm (WPGMA). **B)** Comprison of the assembly-based JAO phylogenetic tree model with the consensual genome-based model. Colors refer to the predefined cladal groups, as in Figures 4, and indicate the cladal cohesiveness of the clustering. Full annotations of the trees are included in Figure 7 and SI.

Closer examination of the resultant tree of life model reveals more reasonable clustering details (**Figure 6A**). Samples that belong to the same species are closely positioned together (see SI), and there is evident logic to the grouping at the lower levels of the tree, such as the cohesive clustering of all Staphylococci species, and the partitioning of halophiles and acidophiles in the archaea domain. Admittedly, several clustering choices specifically at lower taxonomic levels in the model are clearly inaccurate, such as the grouping of nuts and seeds irrespective of their evolutionary associations, and the separate clustering of land and sea animals with the animalia kingdom. This morphological clustering is clearly related to the quality and choice of samples, specifically for multicellular organisms in which separate tissues possess unique metabolic constitutions which may not be representative of the entire organism. Those phenotypic mappings may be overcome by more representative sampling, and by increasing the analyte coverage of the samples, further establishing the unique identities of the molecular fingerprints necessary for classification (see next section). Altogether, it appears that the tree of life model produced by our transformative assembly-based approach, using non-targeted metabolite information, offers insights that appear generally consistent with consensual knowledge on the evolution of species.

*Validation of assembly phylogenetics*

To substantiate our tree of life model, we first inspected the data-driven decisions of our assembly-based phylogenetic algorithm. We observed that samples that are more closely clustered on the tree tend to share a higher degree of high-mass fragments (**Figure 6B**). This can be gleaned from an increasing contribution of low-mass fragments, and a decreasing contribution of high-mass fragments, to the set of fragments shared between more distantly related samples on the tree. This finding substantiates the model and connects it further to the fundamentals of AT, where molecules of higher MAs (which correspond to higher masses (*27*)) appearing across samples indicate an increased degree of shared contingency, as they require more join operations to be constructed. Second, we compared our assembly-based tree to a consensual phylogenetic tree (acquired from the TimeTree database (*38*)), to examine whether our resultant model conform with one constructed from genomic information. We generated distributions of random trees with different levels of constraints and compared them to the genomic tree using the Generalized Robinson-Foulds (GRF) similarity (*39*) (**Figure 6C**) and the Quartet similarity (see SI) as robust phylogenetic analytical metrics (*40*).

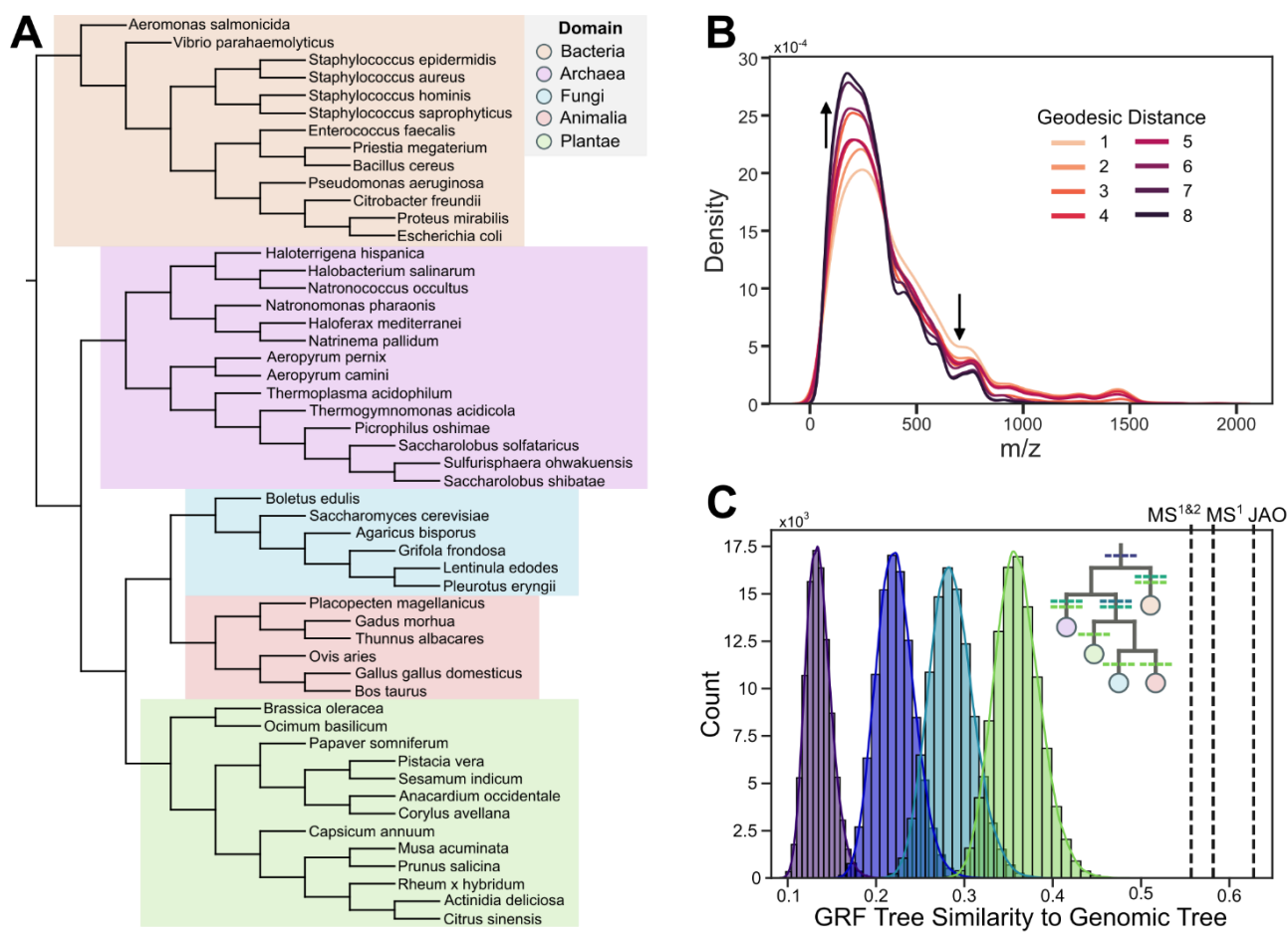

**Figure 6.** Assembly-based phylogenetic Tree of Life from non-targeted metabolite information. **A)** The JAO tree with species annotation. The position of samples within this tree is not pre-determined by *a priori* cladistic knowledge, such as through genomic, proteomic or metabolic information. Colours refer to cladal groups. Branch lengths are uniformly fixed. **B)** Kernel Density Estimate (KDE) plots of the m/z values of all shared $MS^2$ fragments between samples of different geodesic distances in the JAO tree model. Arrows indicate the trends in the data. **C)** Histograms of randomly-generated phylogenetic trees models showing their Generalized Robinson-Foulds (GRF) similarity scores to the consensual genome-based tree model. The different distributions refer to higher degree of constraints in the randomness of the tree generation, depicted in the inset, where randomness was allowed within predefined cladal groups. Purple – completely random, blue – separating prokeryotes and eukaryotes, light blue – separating the bacterial, archaeal and eukaryotic domains, and green – separating the cladal groups of bacteria, archaea, plants, fungi and animals. Colored circles in the inset refer to the cladal groups as appeared in A. The vertical dashed lines represent the score of the phylogenetic tree models, which are 0.58 for $MS^1$ Overlap, 0.56 for $MS^{1\&2}$ Overlap and 0.62 for JAO. T-tests of the JAO model against the random-trees distributions all results in essentially p-value = 0.0.

The GRF similarity metric describes the common taxa between the trees, i.e. the shared sets of samples that are hierarchically clustered in both tree models. The Quartet similarity metric describes the fraction of quartets of samples that are clustered similarly in the tree models. The results of both analyses reveal that all phylogenetic methods produce trees that are substantially non-random and align relatively well with the genomic tree with high statistical significance, and with the assembly-based JAO tree performing better than the controls. Interestingly, the $MS^{1\&2}$ tree presented marginally lower prediction accuracy (with the GRF metric) than the $MS^1$ tree, indicating that the addition of $MS^2$ information adds some interference into the model by treating both types of observations as detected analytes of equal importance. From the data, it appears $MS^2$ information could be used to improve the prediction if processed through the pathway-dependent assembly framework.

As stated above, some clustering choices in our assembly-based phylogenetic tree model appear to be guided by morphology (phenotype) rather than phylogeny, specifically in multicellular organisms where different tissue types present distinct molecular compositions and functions within the organism (**Figure 7A**). This heterogeneity poses a challenge to our inferences, as evolutionary relationships might be masked by common features in samples based on shared phenotypical traits. To address this difficulty, we acquired a new cohort of samples of seven botanical species, from which three types of tissues were sampled – leaves, flowers and branches. We have efficiently processed and analysed the samples together, raising the consistency and quality of the cohort, as well as utilized an automated acquisition protocol that improved the coverage and credibility of the analysis (see methods). This allowed us to detect 25176 unique $MS^1$ analytes and 8028 unique $MS^2$ fragments across the dataset. The resulting tree model presents phylogeny-governed clustering choices mostly aligned with the genomic-based tree (**Figure 7B** and SI). The only contribution of morphological clustering to the tree model occurred at the strain level, indicating that phenotypical features influence the eventual structure of the tree model only when the phylogenetic contribution becomes sufficiently minimal. Interestingly, we noticed some degradation in the processed samples over time, although reanalyses a month and two months later produced highly comparable trees (see

SI), showing phylogenetic modelling to be quite resilient to molecular breakdown. Overall, this experiment emphasizes the importance of sample preparation and quality on credible inferences, which could likely explain clustering errors in our assembly-based preliminary tree of life model.

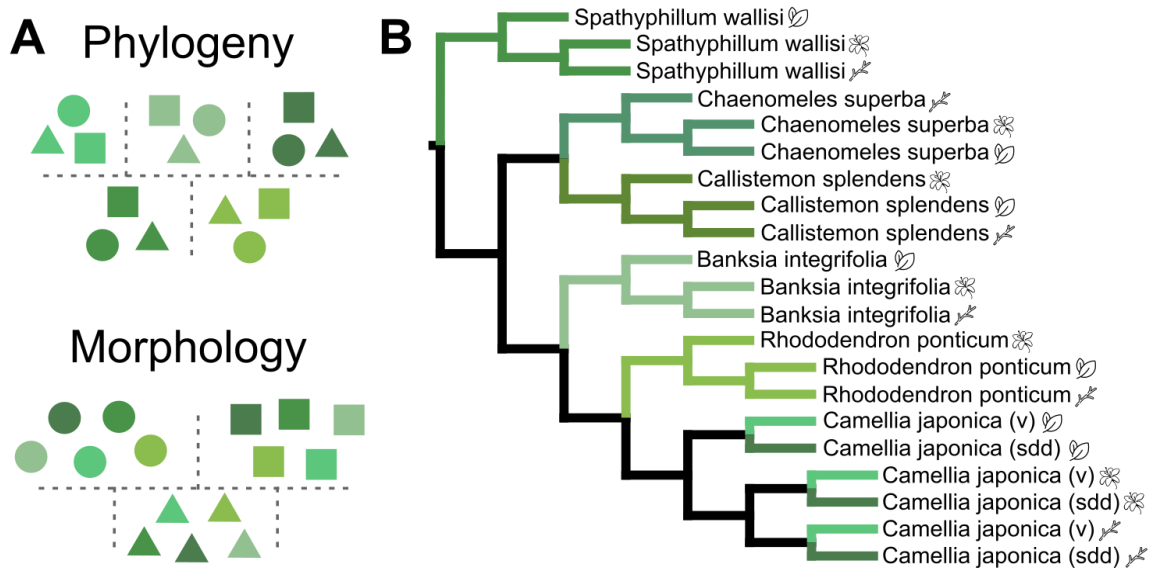

**Figure 7.** Phylogenetic and morphological inferences in multicellular species. **A)** A schematic comparison of clustering choices guided by phylogeny (colour) or morphology (shape). **B)** An assembly-based JAO phylogenetic model of the botanical cohort, consisting of seven species (indicated by colours) and three tissue types (leaves, flowers and branches, specified by icons). Parentheses indicate different strains (v – variegata, sdd – souvenir de desio). The controls produced comparable tree structures (see SI).

*Phylogenetic inference of single-species lineages*

To further substantiate our MS-based phylogenetic approach, we tested it on a case where both morphological and genetic inferences would be challenging and less effective in predicting accurate phylogenetic structures. We decided on a bacterial growth system in which E. coli was recursively cultured and plated, keeping genomic variation to a minimum in the absence of any exerted selection pressure (*41*). In each recursive cycle, an E.coli culture was plated and two colonies were picked from the plate and cultured separately (**Figure 8A**), generating an overall tree-like experimental design comprising four bacterial generations (**Figure 8B**). In this system, the phylogenetic inference was not meant to predict the evolutionary history of distinct species, but that of related colonies of the same species, based on their detected molecular compositions. Therefore this system tests whether assembly-based phylogenetics is able to detect molecular transformations attributed to evolutionary processes on a much shorter timescale than those investigated in a tree of life model, a result of phenotypic variation that is likely not genomically encoded (*42*). In genomics, different regions present diverse mutation rates, and we similarly argue that different metabolomic features vary at disparate rates, and thus an exhaustive analysis of the cellular composome would yield transformations of phylogenetic importance.

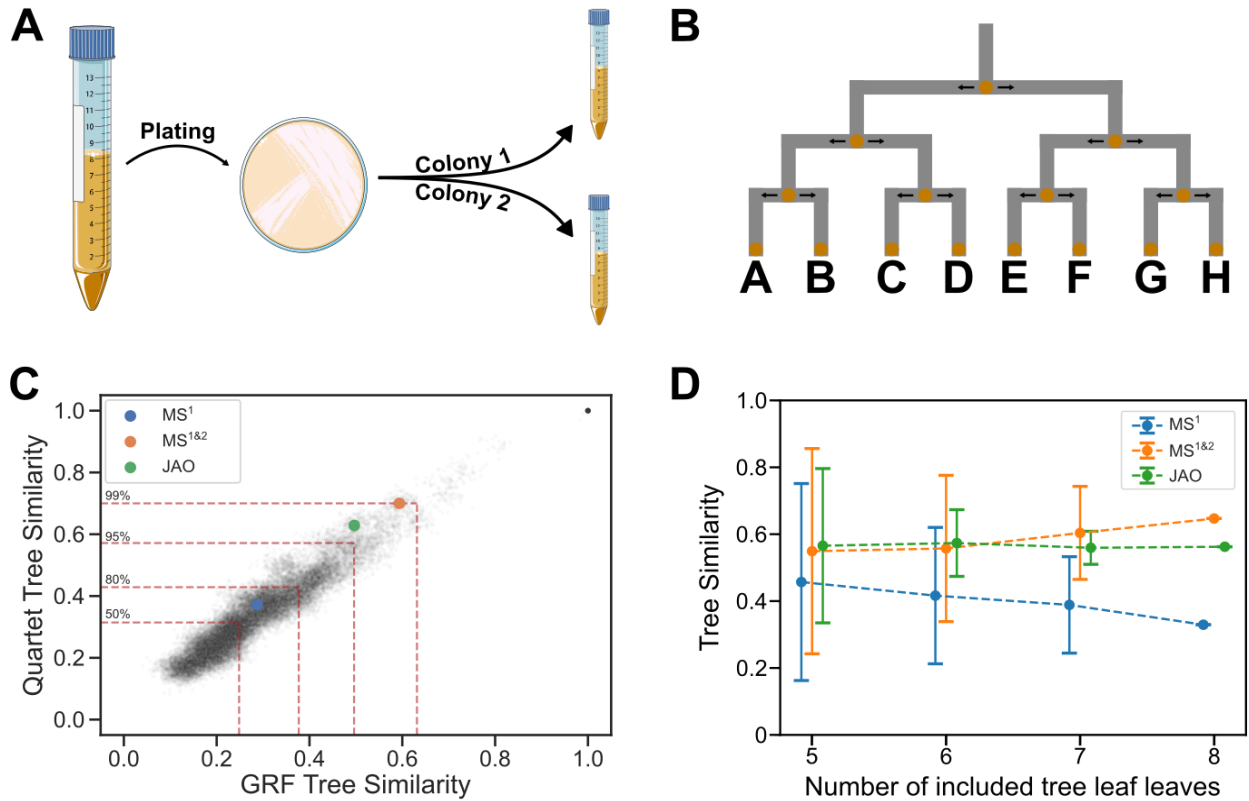

**Figure 8.** Deducing bacterial lineages by Assembly-based phylogenetics. **A)** A schematic representation of a recursive step in the experiment, consisting of plating a culture of E. coli and reculturing two picked colonies from the plate. **B)** The tree-like structure of the recursive experiment, spanning four generations of colony culturing, and resulting in eight leaf node samples indicated by bold letters. **C)** A scatterplot detailing the capacity of the phylogenetic algorithms to predict a tree that closely resembles the actual experimental tree structure, assessed by GRF and Quartet tree similarity metrics. The black dots represent every possible tree structure comprised of eight leaf nodes (135,135 trees), and their tree similarity values with normally-distributed errors for better visual dispersion. The red dashed lines represent percentiles of trees included within the encapsulated areas. **D)** A plot depicting the prediction accuracy of the phylogenetic algorithms in relation to the number of samples included in the tree. The calculations were done through jackknife cross-validation, in which every possible set permutation of n nodes was tested, comparing the resultant predicted tree to the actual experimental tree structure. The tree similarity used was an average of the GRF and Quartet values obtained for each prediction. Error bars represent standard deviations.

The samples analysed revealed a significant degree of homogeneity, as expected. We detected 2186 unique $MS^1$ analytes and 1925 unique $MS^2$ fragments across the eight leaf nodes in the tree, each individual sample comprising about 1400 analytes and 1300 unique fragments. The low variation across the dataset signifies a pronounced difficulty to elucidate the experimental tree structure. Indeed, a simple $MS^1$ Overlap approach gave an almost completely random tree prediction at the 62$^{nd}$ percentile of all possible trees (**Figure 8C**). The assembly-based JAO method performed significantly better, with slightly even better performance for the $MS^{1\&2}$ Overlap method, producing tree predictions at the 96$^{th}$ and 99$^{th}$ percentiles respectively. This means that only a few percentiles of all possible tree structures (a few thousands out of 135,135 possible trees) are closer to the true experimental tree than the assembly-based tree model. This result also indicates that $MS^2$ information could bee more valuable to the detection of evolution on a short timescale than just $MS^1$ analytes. This is further confirmed by the observation that the JAO and $MS^{1\&2}$ Overlap predictions appear to be relatively indifferent to the size of the dataset, whereas the accuracy of the $MS^1$ Overlap approach goes down the more samples are included in the tree (**Figure 8D**). Notably, the assembly-based metric seems to be more consistent and less influenced by outliers than the two control methods, showing lower variance of tree prediction accuracies. These results illustrate the adeptness and reliability of AT to correctly infer phylogenetic and causal relationships from molecular observations in highly homogenous systems that would prove challenging for previously established techniques.

**Conclusions**

The transformative discovery of the genome as a common underlying substrate for recording evolutionary change across biology unified the empirical investigation of all life forms on the planet and enabled the study of deep connections between species in a quantitative, objective manner. With the advent of Assembly Theory, we now have a more general framework within which to study the causal relationships between molecular ensembles of any origin, be it biotic or abiotic. In this work we used this framework to devise a simple quantitative system for detecting and classifying living species without reference to their genetic material, or indeed any prior knowledge of their biochemistry, using a robust general analytical workflow. We presented a non-targeted genome-agnostic empirical examination of molecular compositional information of living and non-living systems and developed models that use this information for accurate phylogenetic inference. We were able to reliably discern the biogenicity of analysed samples and their cladal grouping with supervised and unsupervised methodologies. We then generated phylogenetic tree of life models that closely resemble the consensual genomic tree, all without structure elucidation of molecular components, or any prior genomic, proteomic or metabolic knowledge about the involved samples. To accommodate multicellular organisms, we demonstrated the capacity of our methodology to cluster samples according to phylogeny rather than morphology or function for heterogeneous tissue types. Finally, we were able to track the genealogy of strains of a single bacterial species, strengthening the generality and expediency of our assemble-based approach. These results show that as is the case for genomic analysis, our assembly-based methodology can capture both short-term and long-term evolutionary transformations, and could be employed to map the evolution of life on earth, tracing the contingent history of molecular ensembles without depending on particular chemical or biological features. This approach has the potential to provide an efficient diagnostic tool that may complement genome-based efforts, and should expand our understanding of life, its origin and evolution, beyond current models and definitions that are anchored in the specificity of the biological inception.

**Methods**

*Sample preparation*

A variety of animal, bacteria, fungi, plant, and inorganic samples consisting of complex analyte mixtures were prepared and analysed to garner information on molecular assembly. Animal, plant, and fungi samples were frozen immediately in a -40 °C freezer for 24 hours before freeze-drying (-48 °C, 1.8 mbar) for 48 hours. These were subsequently ground in a mortar and pestle and sieved to produce particle sizes < 0.5 mm. Bacteria samples were cultured overnight in nutrient broth, at 37 °C, then centrifuged to separate the broth supernatant from the bacteria pellet. Inorganic samples were crushed and sieved to produce particle sizes < 0.5 mm. To procure a wide range of metabolites, ultrasound-assisted extraction (UAE, 20 mins/30 °C) was employed with 10 mL of 80% methanol extraction solvent. For each sample, three sample replicates and a control were investigated. Further details of the extraction process are available in the SI section 3. Samples were analysed using HPLC-MS/MS. The comprehensive methodology for the HPLC and Orbitrap Fusion Lumos mass spectrometer are reported in the SI section 4.

*$MS^2$ Optimisation*

The complex nature of our unprocessed samples necessitated an adaptive analysis protocol, implemented via data-dependent MS/MS acquisition (DDA), whereby the most intense analytes from $MS^1$ scans are selected for $MS^2$ fragmentation (*43*). However, DDA can encounter limitations that result in the incomplete sampling of detectible analytes, such as constraints on the number of analytes available for MS/MS due to scan speed, masking of analytes by contamination and noise (*44*). Since it was crucial to improve the abundance and quality of $MS^2$ spectra for these overlooked analytes (*45*), a modified DDA approach was developed whereby an initial MS experiment was conducted to obtain a list of analytes (an inclusion list) unique to each sample, devoid of contamination and noise, before a secondary MS experiment was conducted specifically fragmenting analytes from the inclusion list. Details of the construction of the inclusion list can be found in the SI section 5. For the multi-tissues cohort, as well as for the recursive culturing experiment of E. coli, we progressed to use the AcquireX automated data-acquisition procedure, similar to our modified DDA procedure used for the Tree of Life samples, for improved analyte detection and coverage (*46*).

*Phylogenetic modelling*

Within the generalized AT, we formulated the $MS^1$ Overlap method to be a simple overlap of $MS^1$ analytes between samples, and $MS^{1\&2}$ Overlap to be an overlap of the combined set of $MS^1$ analytes and $MS^2$ fragments between samples. To achieve that, we employed the Otsuka–Ochiai coefficient (OOC) (*37*), a variation of a cosine similarity designed for sets of discrete objects. For two samples A and B, their OOC is calculated thus:

**Equation 1.** $$OOC(A, B) = \frac{|A \cap B|}{\sqrt{|A| \times |B|}}$$

With the generalisation of an assembly space of a single molecule, a corresponding Joint MA index can be calculated for a group molecules. A Joint MA index of a pair of molecules that is lower than the sum of their individual MAs indicates possible shared contingent objects (fragments) in their pathways. Spurred by the Joint MA index's ability to capture the causal relationship between molecules, we proposed a derivation of this metric from tandem MS data of whole samples instead of individual molecules, employing the Recursive MA algorithm (see SI). The Joint Assembly Overlap (JAO) metric could serve as an indicator of a hereditary connection between samples. We formulated JAO of sample A and B to lie between 0 and 1, so that 0 indicates no overlap and 1 indicates full entailment, i.e. all molecules within one sample are contained within the joint assembly space of the other. JAO is calculated thus:

**Equation 2.** $$JAO(A, B) = \frac{JMA(A) + JMA(B)}{JMA(A + B)} - 1$$

with JMA standing for Joint MA.

For a phylogenetic inference, we begun by calculating overlap values for each pair of samples, whether by JAO, $MS^1$ or $MS^{1\&2}$ Overlaps. This is a phenetic approach, producing similarity matrices comprising all samples in the dataset. Following these calculations, we used the Weighted Pair Group Method with Arithmetic Mean (WPGMA) algorithm (implemented with the Scikit-Bio package) to iteratively cluster the clades that exhibit the highest similarity score between them, averaging their similarity scores to all the others groups in the matrix. This procedure is repeated recursively, grouping clades together until reaching a single root clade. All trees were checked with the Mantel test to verify the certainty of tree structure derived from the given data, and for all reported trees the scores were above 0.8.

*Biogenicity determination and cladal group classification*

We performed the classifications using either a supervised leave-one-out cross-validation method or an unsupervised spectral clustering method. In the first method, we iterated over the samples in the dataset, each time one sample was kept aside, and the rest were grouped according to their predefined real groups. We operated over the $MS^1$ Overlap, $MS^{1\&2}$ Overlap and JAO matrices, where the prediction was made based on the maximal values for the examined kept-aside sample with other samples, assigning it to the group of its most-similar pair. For the second method of spectral clustering, we ran the algorithm on the numeric matrices, testing different numbers of clusters, using the *scikit* python package. We reported the Silhouette score for each number of clusters for each metric.

*Structural elucidation of E. coli analytes*

The $MS^2$ files were converted to abf files using AbfConverter (Reifycs). Samples were then checked and searched using MS Dial, an open-source software pipeline (*47*). Samples were searched against the NIST database in .msp format. Tolerances were set at 0.01 Da in the $MS^1$ and 0.05 Da in the $MS^2$. Retention time tolerance was 100 mins, and was not used for scoring or filtering. The matches were then crossed with known E. coli metabolites from a public repository (*48*). In addition, we extracted the SMILES of all metabolites in the repository and in-silico diffracted them using the CFM-IF4 docker software (*49*) to generate ESI MS/MS predicted spectra and compared them to the experimental spectra. Matches were attained if a Jaccard similarity index between the lists of $MS^2$ peaks is above 0.1, and the difference between the masses is below two Daltons.

*Statistical comparison between phylogenetic trees*

We used the R packages TreeDist and Quartet to run the Generalized Robinson-Foulds (GRF) and Quartet similarity calculations, respectively. For the tree of life model, we defined four constraint states for the tree, and for each state generated 100,000 random trees whose randomness followed these constraints. T-tests were then conducted on each of the distributions, which were treated as the null-hypothesis for the assembly tree. For the recursive culturing experiment of E. coli, we generate the 135,135 possible tree constructs that consist of eight leaf nodes, and calculated their GRF and Quartet values to the experimental tree-like structure to anchor the values of the trees predicted by the phylogenetic models. We also conducted a jackknife cross-validation analysis, in which different permutations of samples are tested and the GRF and Quartet scores of the trees generated by the phylogenetic models using these samples are reported.

**Data availability**

All data used to for the analyses given in the SI and the figures will be available on publication on zenodo. For reviewing:

https://www.dropbox.com/scl/fo/bsv3azoqzc9ysvvklolwy/h?rlkey=fwnrz1mcjeza7ud3aezwlt2k7&dl=0

**Author Contributions**

LC conceived the experimental framework building on the concept of the theory with help from SIW. AM developed the original sample operating procedure, and AM and AK collected the samples and retrieved mass spec data with input from EC. AK developed the data analysis approach building the phylogenetic models and the classification methods, and did the statistical analyses. HM expanded the Recursive MA algorithm to emulate joint assembly spaces. MJ and AS helped develop the theoretical framework. LC and AK wrote the manuscript with input from all the authors.

**Acknowledgements**

We would like to thank Michael Lachmann (SFI) for comments on the manuscript, and Keith Patarroyo for his assistance in visualizing assembly pathways. We acknowledge financial support from the John Templeton Foundation (grants 61184 and 62231), EPSRC (grant nos. EP/L023652/1, EP/R01308X/1, EP/S019472/1, and EP/P00153X/1), the Break-through Prize Foundation and NASA (Agnostic Biosignatures award no. 80NSSC18K1140), MINECO (project CTQ2017-87392-P), ERC (project 670467 SMART-POM) and the Alfred P. Sloan Foundation (grant 21110).

**References**

1. E. Mayr, *The Growth of Biological Thought: Diversity, Evolution, and Inheritance* (Harvard University Press, 1982).

2. C. Linnaeus, *Systema Naturae per Regna Tria Naturae, Secundum Classes, Ordines, Genera, Species; Cum Characteribus, Differentiis, Synonymis, Locis* (apud JB Delamolliere, 1789)vol. 1.

3. C. Darwin, W. F. Bynum, *The Origin of Species by Means of Natural Selection: Or, the Preservation of Favored Races in the Struggle for Life* (AL Burt New York, 2009).

4. L. A. Hug, B. J. Baker, K. Anantharaman, C. T. Brown, A. J. Probst, C. J. Castelle, C. N. Butterfield, A. W. Hernsdorf, Y. Amano, K. Ise, Y. Suzuki, N. Dudek, D. A. Relman, K. M. Finstad, R. Amundson, B. C. Thomas, J. F. Banfield, A new view of the tree of life. *Nat. Microbiol.* **1**, 1–6 (2016).

5. F. Delsuc, H. Brinkmann, H. Philippe, Phylogenomics and the reconstruction of the tree of life. *Nat. Rev. Genet.* **6**, 361–375 (2005).

6. S. Mukherjee, R. Seshadri, N. J. Varghese, E. A. Eloe-Fadrosh, J. P. Meier-Kolthoff, M. Göker, R. C. Coates, M. Hadjithomas, G. A. Pavlopoulos, D. Paez-Espino, Y. Yoshikuni, A. Visel, W. B. Whitman, G. M. Garrity, J. A. Eisen, P. Hugenholtz, A. Pati, N. N. Ivanova, T. Woyke, H.-P. Klenk, N. C. Kyrpides, 1,003 reference genomes of bacterial and archaeal isolates expand coverage of the tree of life. *Nat. Biotechnol.* **35**, 676–683 (2017).

7. D. H. Parks, M. Chuvochina, D. W. Waite, C. Rinke, A. Skarshewski, P.-A. Chaumeil, P. Hugenholtz, A standardized bacterial taxonomy based on genome phylogeny substantially revises the tree of life. *Nat. Biotechnol.* **36**, 996–1004 (2018).

8. F. D. Ciccarelli, T. Doerks, C. Von Mering, C. J. Creevey, B. Snel, P. Bork, Toward Automatic Reconstruction of a Highly Resolved Tree of Life. *Science* **311**, 1283–1287 (2006).

9. A. E. Gorbalenya, M. Krupovic, A. Mushegian, A. M. Kropinski, S. G. Siddell, A. Varsani, M. J. Adams, A. J. Davison, B. E. Dutilh, B. Harrach, R. L. Harrison, S. Junglen, A. M. Q. King, N. J. Knowles, E. J. Lefkowitz, M. L. Nibert, L. Rubino, S. Sabanadzovic, H. Sanfaçon, P. Simmonds, P. J. Walker, F. M. Zerbini, J. H. Kuhn, International Committee on Taxonomy of Viruses Executive Committee, The new scope of virus taxonomy: partitioning the virosphere into 15 hierarchical ranks. *Nat. Microbiol.* **5**, 668–674 (2020).

10. C. L. Mills, P. J. Beuning, M. J. Ondrechen, Biochemical functional predictions for protein structures of unknown or uncertain function. *Comput. Struct. Biotechnol. J.* **13**, 182–191 (2015).

11. D. H. Parks, C. Rinke, M. Chuvochina, P.-A. Chaumeil, B. J. Woodcroft, P. N. Evans, P. Hugenholtz, G. W. Tyson, Recovery of nearly 8,000 metagenome-assembled genomes substantially expands the tree of life. *Nat. Microbiol.* **2**, 1533–1542 (2017).

12. P. Kapli, Z. Yang, M. J. Telford, Phylogenetic tree building in the genomic age. *Nat. Rev. Genet.* **21**, 428–444 (2020).

13. K. B. Muchowska, S. J. Varma, J. Moran, Nonenzymatic Metabolic Reactions and Life's Origins. *Chem. Rev.* **120**, 7708–7744 (2020).

14. D. Lancet, R. Zidovetzki, O. Markovitch, Systems protobiology: origin of life in lipid catalytic networks. *J. R. Soc. Interface* **15**, 20180159 (2018).

15. J. C. Bowman, A. S. Petrov, M. Frenkel-Pinter, P. I. Penev, L. D. Williams, Root of the Tree: The Significance, Evolution, and Origins of the Ribosome. *Chem. Rev.* **120**, 4848–4878 (2020).

16. N. Singhal, M. Kumar, P. K. Kanaujia, J. S. Virdi, MALDI-TOF mass spectrometry: an emerging technology for microbial identification and diagnosis. *Front. Microbiol.* **6** (2015).

17. S. Sauer, M. Kliem, Mass spectrometry tools for the classification and identification of bacteria. *Nat. Rev. Microbiol.* **8**, 74–82 (2010).

18. A. T. L. Lun, K. Swaminathan, J. W. H. Wong, K. M. Downard, Mass Trees: A New Phylogenetic Approach and Algorithm to Chart Evolutionary History with Mass Spectrometry. *Anal. Chem.* **85**, 5475–5482 (2013).

19. K. M. Downard, Sequence-Free Phylogenetics with Mass Spectrometry. *Mass Spectrom. Rev.* **41**, 3–14 (2022).

20. S. Han, W. Van Treuren, C. R. Fischer, B. D. Merrill, B. C. DeFelice, J. M. Sanchez, S. K. Higginbottom, L. Guthrie, L. A. Fall, D. Dodd, M. A. Fischbach, J. L. Sonnenburg, A metabolomics pipeline for the mechanistic interrogation of the gut microbiome. *Nature* **595**, 415–420 (2021).

21. T. R. Sandrin, J. E. Goldstein, S. Schumaker, MALDI TOF MS profiling of bacteria at the strain level: A review. *Mass Spectrom. Rev.* **32**, 188–217 (2013).

22. R. A. Musah, E. O. Espinoza, R. B. Cody, A. D. Lesiak, E. D. Christensen, H. E. Moore, S. Maleknia, F. P. Drijfhout, A High Throughput Ambient Mass Spectrometric Approach to Species Identification and Classification from Chemical Fingerprint Signatures. *Sci. Rep.* **5**, 11520 (2015).

23. J. Chalupová, M. Raus, M. Sedlářová, M. Šebela, Identification of fungal microorganisms by MALDI-TOF mass spectrometry. *Biotechnol. Adv.* **32**, 230–241 (2014).

24. R. Gregor, M. Probst, S. Eyal, A. Aksenov, G. Sasson, I. Horovitz, P. C. Dorrestein, M. M. Meijler, I. Mizrahi, Mammalian gut metabolomes mirror microbiome composition and host phylogeny. *ISME J.* **16**, 1262–1274 (2022).

25. C. H. Johnson, J. Ivanisevic, G. Siuzdak, Metabolomics: beyond biomarkers and towards mechanisms. *Nat. Rev. Mol. Cell Biol.* **17**, 451–459 (2016).

26. R. D'Ari, J. Casadesús, Underground metabolism. *BioEssays* **20**, 181–186 (1998).

27. S. M. Marshall, C. Mathis, E. Carrick, G. Keenan, G. J. T. Cooper, H. Graham, M. Craven, P. S. Gromski, D. G. Moore, Sara. I. Walker, L. Cronin, Identifying molecules as biosignatures with assembly theory and mass spectrometry. *Nat. Commun.* **12**, 3033 (2021).

28. M. Jirasek, A. Sharma, J. R. Bame, S. H. M. Mehr, N. Bell, S. M. Marshall, C. Mathis, A. Macleod, G. J. T. Cooper, M. Swart, R. Mollfulleda, L. Cronin, Determining Molecular Complexity using Assembly Theory and Spectroscopy. arXiv arXiv:2302.13753 [Preprint] (2023). https://doi.org/10.48550/arXiv.2302.13753.

29. Y. Liu, C. Mathis, M. D. Bajczyk, S. M. Marshall, L. Wilbraham, L. Cronin, Exploring and mapping chemical space with molecular assembly trees. *Sci. Adv.* **7**, eabj2465 (2021).

30. A. Sharma, D. Czégel, M. Lachmann, C. P. Kempes, S. I. Walker, L. Cronin, Assembly theory explains and quantifies selection and evolution. *Nature*, 1–8 (2023).

31. D. Radoš, S. Donati, M. Lempp, J. Rapp, H. Link, Homeostasis of the biosynthetic E. coli metabolome. *Iscience* **25** (2022).

32. A. Rosato, L. Tenori, M. Cascante, P. R. De Atauri Carulla, V. A. Martins dos Santos, E. Saccenti, From correlation to causation: analysis of metabolomics data using systems biology approaches. *Metabolomics* **14**, 1–20 (2018).

33. E. Borenstein, M. Kupiec, M. W. Feldman, E. Ruppin, Large-scale reconstruction and phylogenetic analysis of metabolic environments. *Proc. Natl. Acad. Sci.* **105**, 14482–14487 (2008).

34. R. Cilibrasi, P. M. B. Vitanyi, Clustering by compression. *IEEE Trans. Inf. Theory* **51**, 1523–1545 (2005).

35. M. Sindelar, G. J. Patti, Chemical Discovery in the Era of Metabolomics. *J. Am. Chem. Soc.* **142**, 9097–9105 (2020).

36. J.-L. Reymond, L. Ruddigkeit, L. Blum, R. van Deursen, The enumeration of chemical space. *WIREs Comput. Mol. Sci.* **2**, 717–733 (2012).

37. V. Verma, R. K. Aggarwal, A comparative analysis of similarity measures akin to the Jaccard index in collaborative recommendations: empirical and theoretical perspective. *Soc. Netw. Anal. Min.* **10**, 43 (2020).

38. S. Kumar, M. Suleski, J. M. Craig, A. E. Kasprowicz, M. Sanderford, M. Li, G. Stecher, S. B. Hedges, TimeTree 5: an expanded resource for species divergence times. *Mol. Biol. Evol.* **39**, msac174 (2022).

39. M. R. Smith, Information theoretic generalized Robinson–Foulds metrics for comparing phylogenetic trees. *Bioinformatics* **36**, 5007–5013 (2020).

40. M. R. Smith, Robust analysis of phylogenetic tree space. *Syst. Biol.* **71**, 1255–1270 (2022).

41. H. Lee, E. Popodi, H. Tang, P. L. Foster, Rate and molecular spectrum of spontaneous mutations in the bacterium Escherichia coli as determined by whole-genome sequencing. *Proc. Natl. Acad. Sci.* **109**, E2774–E2783 (2012).

42. S. Ecker, V. Pancaldi, A. Valencia, S. Beck, D. S. Paul, Epigenetic and Transcriptional Variability Shape Phenotypic Plasticity. *BioEssays* **40**, 1700148 (2018).

43. C. D. Broeckling, E. Hoyes, K. Richardson, J. M. Brown, J. E. Prenni, Comprehensive Tandem-Mass-Spectrometry Coverage of Complex Samples Enabled by Data-Set-Dependent Acquisition. *Anal. Chem.* **90**, 8020–8027 (2018).

44. J. P. Koelmel, N. M. Kroeger, E. L. Gill, C. Z. Ulmer, J. A. Bowden, R. E. Patterson, R. A. Yost, T. J. Garrett, Expanding Lipidome Coverage Using LC-MS/MS Data-Dependent Acquisition with Automated Exclusion List Generation. *J. Am. Soc. Mass Spectrom.* **28**, 908–917 (2017).

45. V. Davies, J. Wandy, S. Weidt, J. J. J. van der Hooft, A. Miller, R. Daly, S. Rogers, Rapid Development of Improved Data-Dependent Acquisition Strategies. *Anal. Chem.* **93**, 5676–5683 (2021).

46. B. Cooper, R. Yang, An assessment of AcquireX and Compound Discoverer software 3.3 for non-targeted metabolomics. *Sci. Rep.* **14**, 4841 (2024).

47. H. Tsugawa, T. Cajka, T. Kind, Y. Ma, B. Higgins, K. Ikeda, M. Kanazawa, J. VanderGheynst, O. Fiehn, M. Arita, MS-DIAL: data-independent MS/MS deconvolution for comprehensive metabolome analysis. *Nat. Methods* **12**, 523–526 (2015).

48. A. C. Guo, T. Jewison, M. Wilson, Y. Liu, C. Knox, Y. Djoumbou, P. Lo, R. Mandal, R. Krishnamurthy, D. S. Wishart, ECMDB: the E. coli Metabolome Database. *Nucleic Acids Res.* **41**, D625–D630 (2012).

49. F. Wang, J. Liigand, S. Tian, D. Arndt, R. Greiner, D. S. Wishart, CFM-ID 4.0: More Accurate ESI-MS/MS Spectral Prediction and Compound Identification. *Anal. Chem.* **93**, 11692–11700 (2021).